\documentclass[preprint, superscriptaddress, amsmath, amssymb, aip, apl]{revtex4-1}

\usepackage{graphicx}
\usepackage{hyperref}
\usepackage{xcolor}
\hypersetup{
    colorlinks=true,
    linkcolor={blue!50!black},
    citecolor={blue!50!black},
    urlcolor={blue!80!black}
}

\makeatletter

\def\@fnsymbol#1{\ensuremath{\ifcase#1\or *\or *\or *\or
   *\or *\or * \or * \or *
   \or \ddagger\ddagger \else\@ctrerr\fi}}

\makeatother

\setcitestyle{super}

\begin{document}

\title{Tuning of a skyrmion cluster in magnetoelectric Cu$_2$OSeO$_3$ by electric field}

\author{\surname{Huang} Ping}
\email{ping.huang@epfl.ch}
\affiliation{Laboratory for Quantum Magnetism (LQM), Institute of Physics, \'{E}cole Polytechnique F\'{e}d\'{e}rale de Lausanne (EPFL), CH-1015 Lausanne, Switzerland}
\affiliation{Laboratory for Ultrafast Microscopy and Electron Scattering (LUMES), Institute of Physics, \'{E}cole Polytechnique F\'{e}d\'{e}rale de Lausanne (EPFL), CH-1015 Lausanne, Switzerland}

\author{Marco Cantoni}
\affiliation{Centre Interdisciplinaire de Microscopie \'{E}lectronique (CIME), \'{E}cole Polytechnique F\'{e}d\'{e}rale de Lausanne (EPFL), CH-1015 Lausanne, Switzerland}

\author{Arnaud Magrez}
\affiliation{Crystal Growth Facility, Institute of Physics,  \'{E}cole Polytechnique F\'{e}d\'{e}rale de Lausanne (EPFL), CH-1015 Lausanne, Switzerland}

\author{Fabrizio Carbone}
\affiliation{Laboratory for Ultrafast Microscopy and Electron Scattering (LUMES), Institute of Physics, \'{E}cole Polytechnique F\'{e}d\'{e}rale de Lausanne (EPFL), CH-1015 Lausanne, Switzerland}

\author{Henrik M. R{\o}nnow}
\affiliation{Laboratory for Quantum Magnetism (LQM), Institute of Physics, \'{E}cole Polytechnique F\'{e}d\'{e}rale de Lausanne (EPFL), CH-1015 Lausanne, Switzerland}

\begin{abstract}

Chiral magnetic textures with non-trivial topology are known as skyrmions, and due to their unique properties they are promising in novel magnetic storage applications. While the electric manipulation of either isolated skyrmions or a whole skyrmion lattice have been intensively reported, the electric effects on skyrmion clusters remain scarce. In magnetoelectric compound Cu$_2$OSeO$_3$, a skyrmion cluster can be created near the helical-skyrmion phase boundary. Here, we report the in situ electric field writing/erasing of skyrmions in such a skyrmion cluster. Our real space/time image data obtained by Lorentz transmission electron microscopy and the quantitative analysis evidence the linear increase of the number of skyrmions in the cluster upon the application of a creating electric field. The energy needed to create a single skyrmion is estimated to be $\mathcal{E}=4.7 \times 10^{-24}$ J.

\end{abstract}

\date{\today}

\maketitle

\section*{Introduction}

Skyrmions are vortex-like spin textures with non-trivial topology\cite{muhlbauer_skyrmion_2009, yu_real-space_2010}. As localized objects with nano-metric dimensions and the ability of motion driven by various excitations\cite{nagaosa_topological_2013}, they are very promising in the next generation spintronic applications\cite{fert_skyrmions_2013}.

One of the most attractive features of skyrmions is their ability of being manipulated via divergent magnetic or electric effects, including electric current induced motions\cite{yu_skyrmion_2012, jiang_blowing_2015} and rotations\cite{jonietz_spin_2010}, electric field induced creation\cite{hsu_electric-field-driven_2016, huang_situ_2018} and rotations\cite{white_electric-field-induced_2014}, and magnetic field induced skyrmion phase transitions\cite{rajeswari_filming_2015, huang_melting_2018}. However, most of these effects were mainly realized in systems consisting of either isolated skyrmions or a two-dimension triangular skyrmion lattice, investigations on clusters of skyrmions\cite{muller_magnetic_2017}, on the other hand, remain lacking. In their potential application, e.g. in racetrack magnetic storage, skyrmions are mainly in the form of clusters\cite{fert_skyrmions_2013}, thus the controllable creation and manipulation of skyrmion clusters are pivotal.

In bulk skyrmion systems such as $B$20 compounds Cu$_2$OSeO$_3$, especially in a nano-slab sample, a skyrmion cluster can be created near the phase boundaries between the skyrmion lattice phase and the helical phase\cite{muller_magnetic_2017}. This is due to the fact that the helical-skyrmion transition is a first order phase transition so that phase coexistence happens near the phase boundary, as schematically illustrated in Fig.~\ref{fig:fig1}b. More interestingly, Cu$_2$OSeO$_3$ exhibits complicated magnetoelectric coupling behaviors\cite{seki_magnetoelectric_2012, white_electric-field-induced_2014, huang_situ_2018}, which provides a unique handle of tuning the magnetic textures via electric field\cite{okamura_transition_2016, kruchkov_direct_2018, huang_situ_2018} and vise versa. Here we show that skyrmions can be written and erased in such a skyrmion cluster by electric field ($E$-field). Through the in situ Lorentz transmission electron microscopy (LTEM) investigation, we show directly that one polarity of the applied $E$-field enlarges the skyrmion cluster while the opposite one shrinks it due to the $E$-field induced decrease and increase of the skyrmion free energy respectively. The quantitative statistics of the number of skyrmions from image data reveals a linear dependence of the number of skyrmions on the applied $E$-field, from which the creation energy can be estimated to be $\mathcal{E}=4.7 \times 10^{-24}$ J per skyrmion.

\section*{Sample and experiments}

\begin{figure*}[!htb]
  \centering
  \includegraphics{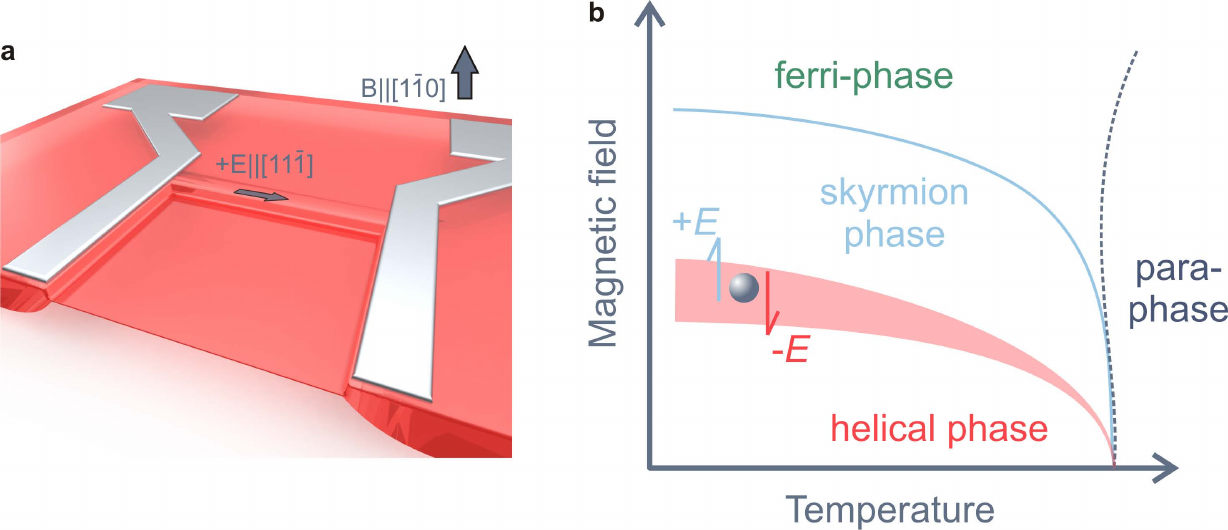}
  \caption{\textbf{a.} Illustration of the sample configuration. Two Pt electrodes were directly deposited by FIB onto the upper surface of the sample. A 150 nm thick TEM lamella was fabricated by FIB at the edge parallel to the $[11\bar{1}]$ direction, along which the in-plane $E$-field is applied. The perpendicular magnetic field is along the $[1\bar{1}0]$ direction. \textbf{b.} Schematic phase diagram indicating the concept and conditions of the experiment. Below the magnetic ordering temperature 60 K, a nano-slab Cu$_2$OSeO$_3$ sample possesses the helical spin ground state. Increasing the magnetic field results in a skyrmion phase. Due to the first order nature of the phase transition, there is helical-skyrmion coexisting regime, as denoted by the shaded region. The sample was initially set inside this region so that a skyrmion cluster surrounded by spin helices can be observed. The application of an $E$-field will then tune the free energy of the system so as to tune the size of the skyrmion cluster, as indicated by the arrows.}
  \label{fig:fig1}
\end{figure*}

A single crystal of Cu$_2$OSeO$_3$ was aligned by X-ray Laue diffraction and cut into a rectangularly shaped slice with the dimensions of 1.5$\times$1.0$\times$0.5 mm$^3$. The main plane is perpendicular to [1$\bar{1}$0] and the two edges are along [11$\bar{1}$] and [112] respectively. In such a geometry, when applying an out-of-plane magnetic field along [1$\bar{1}0$], spontaneous electric polarization will emerge along [00$\bar{1}$] in the skyrmion phase\cite{seki_magnetoelectric_2012, huang_situ_2018}. The slice was mechanically polished down to 20 $\mu$m thick and a TEM lamella of 22 $\mu$m wide was further milled to 150 nm by focused ion beam (FIB) to form the H-bar sample geometry. Two Pt electrodes 57 $\mu$m separated were deposited by FIB directly on the top surface of the sample, and a 7 nm amorphous carbon layer was deposited on the bottom surface to avoid charging effect. The sample configuration is illustrated in Fig.~\ref{fig:fig1}

The sample was mounted on a liquid helium TEM sample holder (Gatan), and Low temperature vacuum grease was used to ensure the thermal contact. The two Pt electrodes were connected to the electric feed-throughs of the sample holder via cooper wires by silver paste. A Keithley 2400 source-meter was used to apply the electric fields while the leakage currents were monitored and recorded simultaneously. The highest applied voltage was $|U|=$ 200 V, corresponding to an $E$-field of $|E|=$3.6 V/$\mu$m.

As schematically illustrated in Fig.~\ref{fig:fig1}b, Cu$_2$OSeO$_3$ forms magnetically ordered states below the ordering temperature 60 K, and the magnetic ground state is spin helices with multiple propagation vectors. For a nano-slab Cu$_2$OSeO$_3$ sample, a skyrmion phase emerges with the increase of the magnetic field, and this phase spans to the base temperatures, in contrast to the case in bulk samples in which the skyrmion phase is squeezed to a tiny portion of the phase diagram by a magnetic field polarized spin conical phase\cite{seki_observation_2012, muhlbauer_skyrmion_2009}. Further increase of the magnetic field results in a ferri-magnetic phase in the 3-up-1-down configuration\cite{bos_magnetoelectric_2008}.

The sample was cooled down in zero magnetic field to 24.7 K to achieve the helical spin ground state. Then the magnetic field was carefully tuned so that a cluster of skyrmions appeared in the helical phase background, as shown in Fig.~\ref{fig:fig2}a. The resulting magnetic field was $H=408$ Oe. A sequence of electric field was then applied, as shown in Fig.~\ref{fig:fig3}b, and a LTEM  movie was acquired simultaneously with the frame interval of 250 ms. The magnetic field and the temperature were both kept constant throughout the whole process of data acquisition.

\section*{Results and analysis}

\begin{figure*}[!htb]
  \centering
  \includegraphics{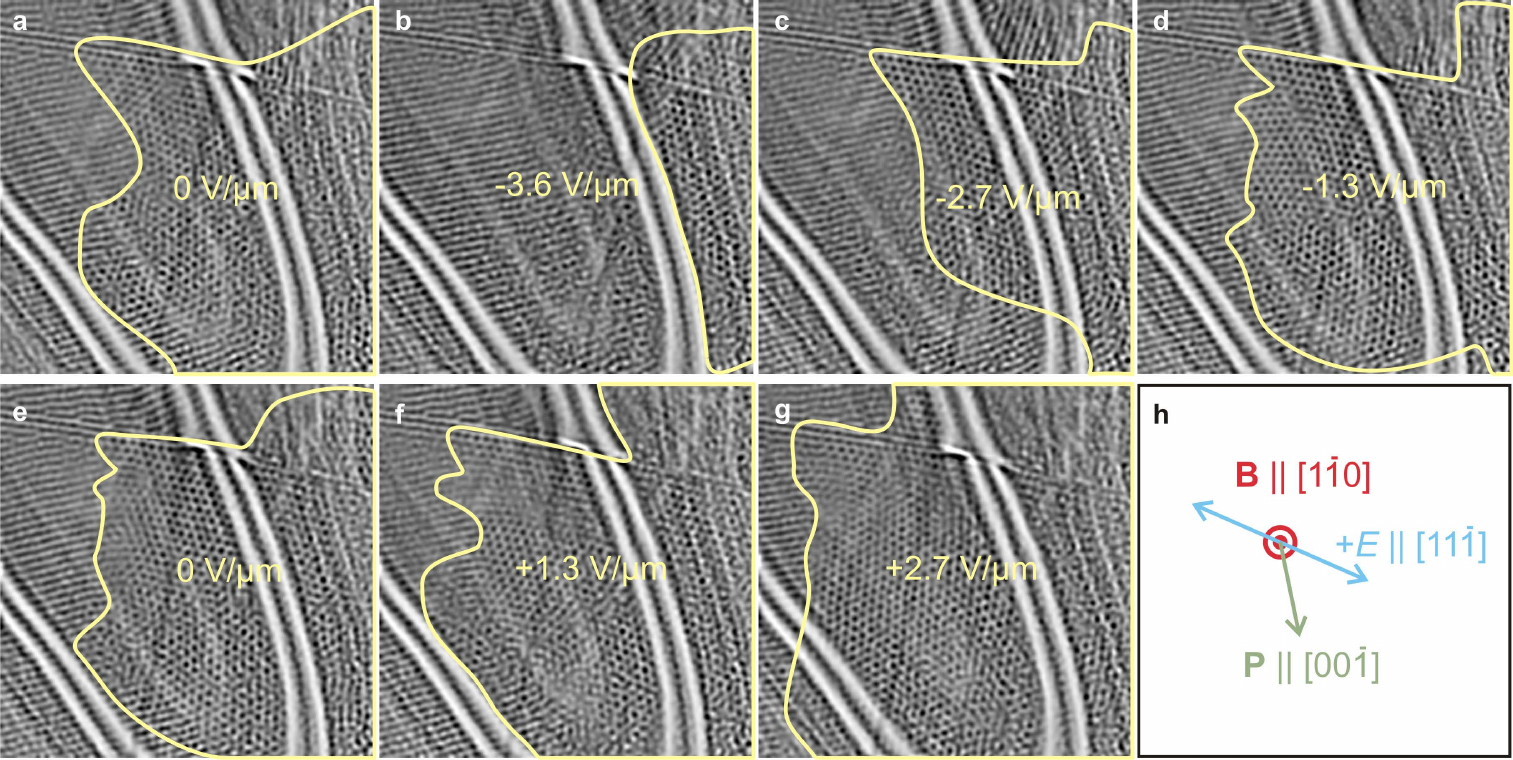}
  \caption{\textbf{a.} Real space image of the initial state of the system after zero-field cooling down to $T=$ 24.7 K and then applying $H=$ 408 Oe. A skyrmion cluster can be seen embedded in the helical phase background. \textbf{b.} to \textbf{g.} Real space LTEM images after applying a sequence of $E$-fields. The size of the initially created skyrmion cluster is observed to be efficiently tuned by the applied $E$-field. The skyrmion clusters are highlighted and the values of the $E$-fields are labeled in each panel. \textbf{h.} The experimental configuration with respect to the images. Note that the coupling between the positive $E$-field $+E$ and the emerged polarization $\mathbf{P}$ lowers down the free energy of the skyrmion phase and the coupling between $-E$ and $\mathbf{P}$ increase that.}
  \label{fig:fig2}
\end{figure*}

Fig.~\ref{fig:fig2} b to g show the LTEM images at different applied $E$-fields, as labeled in each panel. The application of the highest negative $E$-field shrinks the skyrmion cluster dramatically compared to the original state in Fig.~\ref{fig:fig2}a. The size of the skyrmion cluster recovers with the decrease of the amplitude of the negative $E$-field, and almost the same size can be obtained when the $E$-field goes back to 0 V/$\mu$m (Fig.~\ref{fig:fig2}e). The application of positive $E$-fields is observed to further enlarge the size of the skyrmion cluster.

These results demonstrate the $E$-field effect on skyrmions in such a configuration: positive $E$-fields write skyrmions and negative ones erase them. Fig.~\ref{fig:fig2}h indicates the directions of the external $E$-fields and the spontaneous skyrmion polarizations respectively with respect to the real space images. Due to the coupling between the $E$-field and the polarization $\mathcal{E} \propto -\mathbf{E} \cdot \mathbf{P}$, positive $E$-fields decrease and negative ones increase the free energy of the skyrmion phase, so that skyrmions can be created and annihilated correspondingly\cite{kruchkov_direct_2018}.

Deeper understanding of our results can be achieved by quantitative statistics on the number of skyrmions from the image data. To do this, we developed an algorithm that combines together the minimum searching and the orientational intensity distribution calculation\cite{rao_computing_1989, rao_computerized_1992}, so that good balance between precision and efficiency can be obtained\cite{huang_situ_2018}.

\begin{figure*}[!htb]
  \centering
  \includegraphics{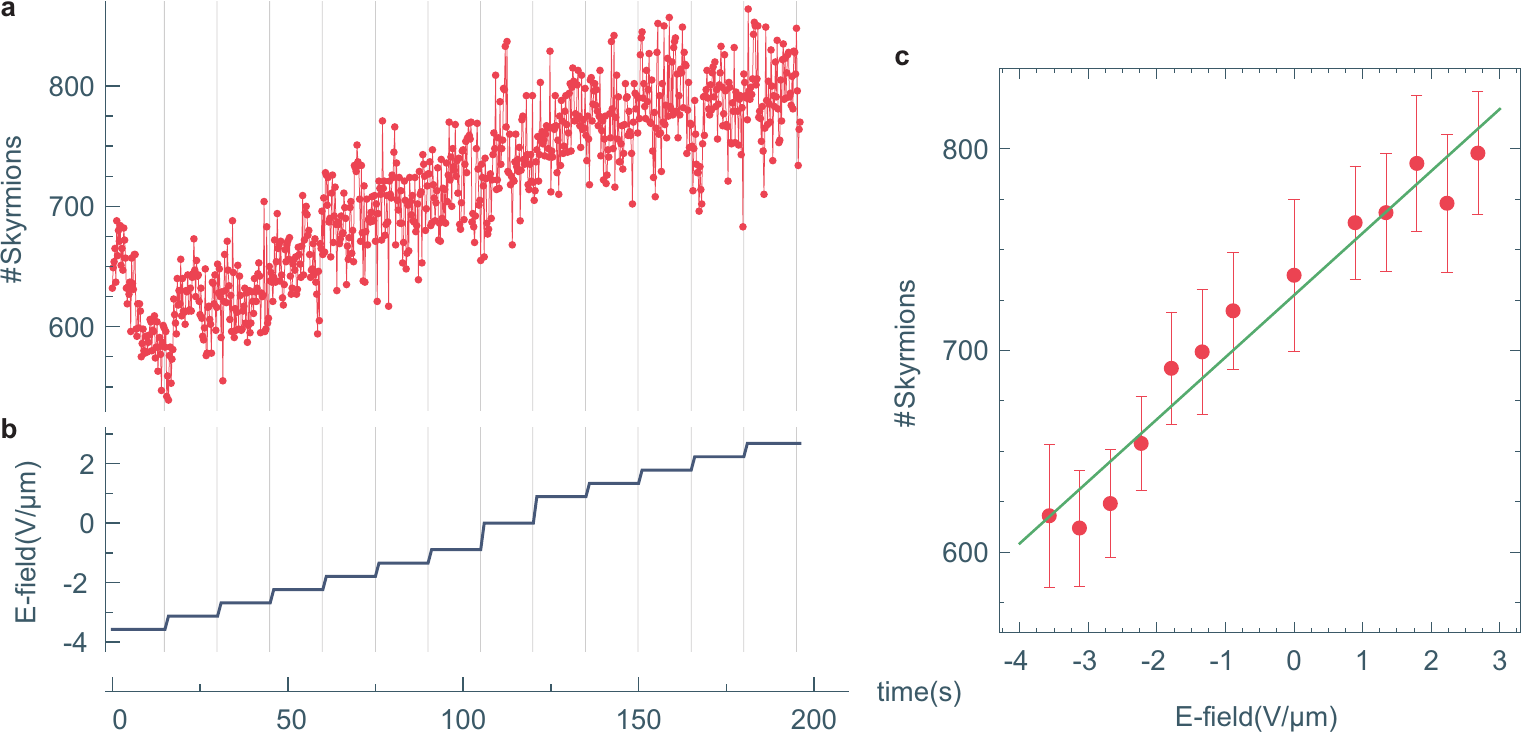}
  \caption{\textbf{a.} The counted number of skyrmions by the automatic algorithm in each frame of the LTEM movie as a function of time. \textbf{b.} The corresponding $E$-field as a function of time. The increase of the number of skyrmions as the negative $E$-field decrease or as the positive $E$-field increase can be clearly seen. \textbf{c.} The red dots denote the averaged number of skyrmions within the lasting time of each $E$-field value, which can be well fitted linearly, as indicated by the green line.}
  \label{fig:fig3}
\end{figure*}

The counting results are shown in Fig.~\ref{fig:fig3}a. Initially, finite numbers of skyrmions are revealed at $E=$0 V/$\mu$m. Then a sharp decrease of the number of skyrmions can be observed after applying the highest negative $E$-field, $E=-$3.6 V/$\mu$m. Increase of the number of skyrmions corresponds perfectly with the change of the $E$-field towards the positive side, reaching its peak at the largest positive $E$-field, $E=+$3.6 V/$\mu$m. This phenomenon is fully reversible, i.e. changing back of the electric field to the negative side will again suppress the number of skyrmions (data not shown here).

The counted numbers of skyrmions at each $E$-field step are averaged and the results are shown in Fig.~\ref{fig:fig3}c. The $E$-field dependence of the number of skyrmions can be well fitted linearly, revealing a slope of $k= dN_{sk}/dE = $ 34 skyrmions per unit $E$-field (V/$\mu$m), as indicated by the green line in Fig.~\ref{fig:fig3}c. Given the polarization $P=0.5$ $\mu$C/m$^2$ for the skyrmion phase of Cu$_2$OSeO$_3$\cite{seki_magnetoelectric_2012}, and the dimension of a single skyrmion unit cell in our sample $S = a \cdot \frac{\sqrt{3}a}{2}$ ($a=65$ nm is the skyrmion lattice constant derived from our image data and is consistent with the reported value\cite{seki_observation_2012}), the energy needed to create a single skyrmion can be estimated as:

\begin{eqnarray}
 \begin{aligned}
  \mathcal{E} &= \frac{\mathbf{E} \cdot \mathbf{p}}{N_{sk}} \\
              &= \frac{EPV\cos{\theta}}{N_{sk}} \\
              &= \frac{PV\cos{\theta}}{k}
 \end{aligned}
\nonumber
\end{eqnarray}

where $\theta$ is the angle between the $E$-field and the polarization with $\cos{\theta}=1/\sqrt{3}$, and $V=St$ is the volume of a single unit cell of the skyrmion lattice ($t=$150 nm is the thickness of the sample). This estimate results in a creation energy per skyrmion to be $\mathcal{E}=4.7 \times 10^{-24}$ J. This value is much smaller than the estimate in our previous work, in which the $E$-field creation of skyrmion directly from the helical phase was described\cite{huang_situ_2018}. Note that the counting of skyrmions in the latter work was carried out in a limit field of view, thus there is a possibility of exaggerating the average energy due to the underestimating of the number of skyrmions outside the field of view. While in the current work, the energy is estimated via the rate of change of the number of skyrmion upon the application of an $E$-field, thus the estimation error may be reduced. Since LTEM is a rather local technique, the difference in the creation energy may also be induced by the different local conditions.

In summary, we realized the in situ $E$-field writing/erasing of skyrmions in a skyrmion cluster in the helical spin background of Cu$_2$OSeO$_3$ via real space/time LTEM investigation. Our quantitative analysis on the real space LTEM movie reveals that this $E$-field approach is reversible and efficient, which implies its potential in skyrmion based spintronic applications.

\acknowledgments
This work was supported by the Swiss National Science Foundation (SNSF) through project 166298, the Sinergia network 171003 for Nanoskyrmionics, and the National Center for Competence in Research 157956 on Molecular Ultrafast Science and Technology (NCCR MUST), as well as the ERC project CONQUEST.

\end{document}